\documentstyle[prd,aps,preprint]{revtex}
\tightenlines
\input epsf.tex
\def\DESepsf(#1 width #2){\epsfxsize=#2 \epsfbox{#1}}
\def\Journal#1#2#3#4{{#1} {\bf #2}, #3 (#4)}

\def\NPB{{\em Nucl. Phys.} B}
\def\PLB{{\em Phys. Lett.}  B}
\def\PRL{\em Phys. Rev. Lett.}
\def\PRD{{\em Phys. Rev.} D}
\def\ZPC{{\em Z. Phys.} C}

\begin{document}

\draft
\preprint{\hbox{CTP-TAMU-08-01}}
\title{Muon $g-2$, Dark  Matter Detection \\ 
and  Accelerator Physics} 

\author{R. Arnowitt, B. Dutta, B. Hu and Y. Santoso }

\address{
Center For Theoretical Physics, Department of Physics, Texas A$\&$M
University, College Station TX 77843-4242}
\date{February, 2001} 
\maketitle
\begin{abstract}
We examine the recently observed deviation of the muon $g - 2$ 
from the Standard Model 
prediction within the framework of gravity mediated SUGRA models  with R parity 
invariance. Universal soft breaking (mSUGRA) models, and models with 
non-universal  Higgs and third generation squark/slepton masses at $M_G$ are 
considered. All relic density constraints from stau-neutralino 
co-annihilation and large $\tan\beta$ NLO corrections for $b \rightarrow
s\gamma$ decay are 
included, and we consider two possibilities for the light Higgs: $m_h > 114$ 
GeV and $m_h > 120$ GeV. The combined $m_h$, $b \rightarrow s\gamma$ and
$a_{\mu}$ bounds give 
rise to lower bounds on $\tan\beta$ and $m_{1/2}$, while the lower bound on
$a_{\mu}$ 
gives rise to an upper bounds on $m_{1/2}$. These bounds are sensitive to $A_0$,
e.g. for $m_h > 114$ GeV, the $95\%$ C.L. is $\tan\beta > 7(5)$ for $A_0 = 0(-4m_{1/2})$, and for
$m_h > 120$ 
GeV, $\tan\beta > 15(10)$. The positive sign of the $a_{\mu}$ deviation implies
$\mu > 
0$, eliminating the extreme cancellations in the dark matter 
neutralino-proton detection cross section so that almost all the SUSY 
parameter space should be accessible to future planned detectors.  Most of 
the allowed parts of parameter space occur in the co-annihilation region 
where $m_0$ is strongly correlated with $m_{1/2}$. The lower bound on $a_{\mu}$ then
greatly 
reduces the allowed parameter space. Thus using $90\%$ C. L. bounds on $a_{\mu}$
we 
find for $A_0 = 0$ that $\tan\beta \geq 10$ and for $\tan\beta \leq 40$ that $m_{1/2}
= (290 - 
550)$ GeV and $m_0 = (70 - 300)$ GeV. Then the tri-lepton signal  and other SUSY 
signals would be beyond the Tevatron Run II (except for the light Higgs), 
only the $\tilde{\tau}_1$ and $h$ and (and for
part of the parameter 
space) the $\tilde{e}_1$ will be accessible to a 500 GeV NLC , while the LHC would be 
able to see the full SUSY mass spectrum. 

\end{abstract}

\newpage

The remarkable accuracy with which the muon gyromagnetic ratio can be 
measured makes it an excellent probe for new physics beyond the Standard 
Model.  The recently reported result of the Brookhaven  E821 experiment now 
gives a $2.6 \sigma$ deviation from the predicted value of the Standard
Model~\cite{b1}:
\begin{equation}
           a_{\mu}^{\rm exp} - a_{\mu}^{\rm SM}  =  43(16) \times 10^{-10}
\end{equation}                                           
where $a_{\mu} = (g_{\mu} - 2)/2$.  Efforts were made initially to calculate a 
possible deviation  from the Standard Model within the framework of global 
supersymmetry (SUSY)~\cite{b2}. However, one may show that in the limit of exact 
global supersymmetry, $a_{\mu}^{\rm SUSY}$ will vanish~\cite{b3}, and thus one
needs broken supersymmetry to obtain a non-zero result. The absence of a 
phenomenologically viable way of spontaneously breaking global 
supersymmetry made realistic predictions for these models difficult. In 
contrast, spontaneous breaking of supersymmetry in supergravity (SUGRA) is 
easy to achieve, and the advent of supergravity grand unified models~\cite{b4} 
led to the first calculations of $a_{\mu}^{\rm SUGRA}$~\cite{b5,b6}, of which
\cite{b6} was
the first complete analysis. Since that time there have been a number of papers 
updating that result. (See e.g. \cite{b7}.)

In SUGRA models,  the spontaneous breaking of supersymmetry triggers the 
Higgs VEV and hence the breaking of $SU(2) \times U(1)$, relating then these two 
mass scales. Thus the scale of the new SUSY masses  is predicted to be 
$\sim 100$ GeV - 1 TeV.  It was then possible to predict in \cite{b6} that the
SUGRA contributions would be comparable or larger than the electroweak 
contribution, $15.2(4) \times 10^{-10}$~\cite{b8}, in accord with the now
observed deviation of Eq. (1). This scale for the SUSY masses was further
confirmed by the LEP 
data showing that consistency with grand unification could be obtained if 
the SUSY masses also lie in the above range~\cite{b9}. Finally, we note that
SUGRA 
models with R-parity invariance predict a dark matter candidate (the 
lightest neutralino) with the astronomically observed amount of relic 
density if the SUSY masses again lie in this range.

It is thus reasonable to investigate whether the observed deviation from 
$a_{\mu}^{\rm exp}$ can be understood within the framework of SUGRA models, and
in this 
paper we consider gravity mediated SUSY breaking with R-parity invariance for
models
with universal soft breaking masses (mSUGRA)  and also models with 
non-universal masses in the Higgs and third generation sector.  SUGRA 
models have a wide range of applicability including cosmological phenomena 
and accelerator physics, and constraints in one area affect predictions in 
other areas.  In particular, as first observed in \cite{b5} and emphasized in
\cite{b10}, that $a_{\mu}$ increases
 with 
$\tan\beta$, as do dark matter detection rates.  Thus as we will see,  the 
deviation of Eq. (1) will significantly effect the minimum neutralino 
-proton cross section, $\sigma_{\tilde{\chi}_1^0-p}$, for terrestial detectors.
Even more significant is the fact that the astronomical
bounds on the $\tilde\chi^0_1$ relic density restrict the SUSY parameter space
and hence the SUGRA predictions for $a_\mu$ as well as what may be expected to
be seen at the Tevatron RUN II and the LHC. In order to 
carry out this analysis, however, it is necessary to  include all the 
co-annihilation effects for large $\tan\beta$, as well as the large $\tan\beta$ 
corrections to $m_b$ and $m_{\tau}$ (which are needed to correctly 
determine the corresponding Yukawa coupling constants) and the large 
$\tan\beta$ NLO corrections to the $b \rightarrow s  \gamma$ decay~\cite{b11}.
In addition, the light Higgs ($h$) mass bounds play an important role in 
limiting the SUSY parameter space and it is necessary to include the one and 
two loop corrections, and the pole mass corrections. The above corrections 
for dark matter (DM) calculations were carried out in \cite{b12}, and we will use 
the same corrections here.  Recently several papers have appeared analysing the
SUGRA contribution to $a_{\mu}$ in light of the final LEP bounds on $m_h$ and
the deviation of Eq.(1)\cite{a,b,b16,d}. Relic density constraints were not
considered in Refs. \cite{b16,d} and coannihilation effects apparently not
included in
Refs. \cite{a,b}. Also Refs. \cite{a,b16,d} do not seem to have included the
constraints from the $b\rightarrow s\gamma$ decay. As will be seen
below, these effects are of major importance in determining the SUGRA
predictions.

Before proceeding on, we state the range of parameters we assume. We take a 
$2 \sigma$ bound of Eq. (1),
\begin{equation}
   11 \times 10^{-10}  <  a_{\mu}^{\rm SUGRA}  <  75 \times 10^{-10}\, ,  
\end{equation}
a $2 \sigma$ bound on the $b \rightarrow s\gamma$ branching ratio,  $1.8 \times
10^{-4} < BR(b \rightarrow s\gamma) < 4.5 \times 10^{-4}$, and a neutralino
relic density range of $ 
0.02  < \Omega_{\tilde{\chi}_1^0} h^2  < 0.25$. (Assuming a lower bound of 0.1
does not 
affect results significantly.)  The $b$-quark mass is assumed to have the 
range $4.0 \,{\rm GeV} < m_b(m_b) < 4.4$ GeV.  We consider two bounds on the
Higgs mass: 
$m_h  > 114$ GeV and $m_h > 120$ GeV. The first is the current LEP bound and 
the second is likely within reach of the Tevatron Run II. However, the 
theoretical calculations of $m_h$ have still some uncertainty as well as 
uncertainty in the $t$-quark mass, and so we will conservatively interpret 
these bounds to mean that our theoretical values obey $m_h > 111$ GeV and 117 
GeV respectively. (Our calculations of $m_h$ are consistent with \cite{b13}.)
The scalar and gaugino masses  at the GUT scale obey $(m_0, 
m_{1/2})  < 1$ TeV. We examine the range  $2 < \tan\beta < 40$, and the cubic
soft 
breaking mass is parameterized  at the GUT scale by $|A_0| < 4 m_{1/2}$. 
Non-universal masses deviate from universality according to $m_0^2( 1 + 
\delta)$ where $-1 < \delta < +1$.  Other parameters are as in \cite{b12}.

We consider first the mSUGRA model, which depends on the four parameters 
$m_0$, $m_{1/2}$,  $A_0$, $\tan\beta = \langle H_2 \rangle / \langle H_1
\rangle $  ( where $\langle H_{(1,2)} \rangle $ give rise to $(d,u)$ 
quark masses) and the sign of the $\mu$ parameter  of the Higgs mixing part of 
the superpotential ($W = \mu H_1 H_2$).  The SUSY contribution to $a_{\mu}$
arises 
from two types of loop diagrams, i. e. those with chargino -sneutrino 
intermediate states, and those with neutralino-smuon intermediate states. 
The dominant contribution arises from the former term with the light 
chargino ($\tilde{\chi}^{\pm}_1$). For moderate or large $\tan\beta$,  and when
($\mu \pm \tilde{m}_2)^2 \ll M_W^2$,  one finds
\begin{equation}
     a_{\mu}^{\rm SUGRA} \cong \frac{\alpha}{4 \pi} \frac{1}{\sin^2 \theta_W}
     \left( \frac{m_{\mu}^2}{m_{\tilde{\chi}^{\pm}_1} \mu} \right) \frac{\tan
     \beta}{1-\frac{\tilde{m}_2^2}{\mu^2}} \left[ 1-\frac{M_W^2}{\mu^2}
    \frac{1+3\frac{\tilde{m}_2^2}{\mu^2}}{\left
(1-\frac{\tilde{m}_2^2}{\mu^2}\right)^2}\right] F(x)
\end{equation}
where $\tilde{m}_i  =(\alpha_i /\alpha_G) m_{1/2}$, $i=1,2,3$ are the gaugino masses
at the electroweak scale and $\alpha_G \cong 1/24$ is the GUT scale gauge
coupling 
constant. (One has $m_{\tilde{\chi}^{\pm}_1} \cong \tilde{m}_2 \cong 0.8
m_{1/2}$, and the gluino ($\tilde{g}$) mass 
is $m_{\tilde{g}} \cong \tilde{m}_3$.) In Eq. (3),  the form factor is $F(x) =
(1 -3x)(1 - x)^{-2} - 2x^2 (1 - x)^{-3}\ln x$, where $x=
(m_{\tilde{\nu}}/m_{\tilde{\chi}^{\pm}})^2$. The sneutrino 
and chargino masses being related to $m_0$ and $m_{1/2}$ by the renormalization 
group equations (RGE)~\cite{b14}. (The contribution from the heavy chargino,
$\tilde\chi^{\pm}_2$ reduces this result by about a third.) One finds for large $m_{1/2}$ that $F(x)
\cong 0.6$
so that  $a_{\mu}$ decreases as $1/m_{1/2}$, while for large $m_0$, $F$
decreases as $\ln(m_0^2)/m_0^2$ (exhibiting the  SUSY decoupling phenomena).

  Eq. (3) exhibits also the fact discussed in \cite{b10,b15} that the sign of 
$a_{\mu}^{\rm SUGRA}$  is given by the sign of $\mu$. Eq. (1) thus implies that
$\mu$ is 
positive  (as pointed out in \cite{b,b16,d}). This then has immediate consequences 
for dark matter detection. Thus as discussed in \cite{b17,b18,b12},  for $\mu <
0$, 
accidental cancellations can occur reducing  the neutralino-proton cross 
section to below $10^{-10}$ pb over a wide range of SUSY parameters,  and 
making halo neutralino dark matter unobservable for present or future 
planned terrestial detectors. Thus this possibility has now been 
eliminated, and future detectors (e.g. GENIUS) should be able to scan almost
the full SUSY parameter space for $m_{1/2} < 1$ TeV.

The lower bound of Eq. (1) plays a central role in limiting the $\mu > 0$ SUSY 
parameter space, particularly when combined with the bounds on the Higgs 
mass and the  $b \rightarrow s\gamma$ constraints. As seen above, lowering 
$\tan\beta$ can be compensated in $a_{\mu}$ by also lowering $m_{1/2}$. However,
$m_h$ 
decreases with both decreasing $\tan\beta$ and decreasing $m_{1/2}$. Thus the 
combined Higgs and $a_{\mu}$ bounds put a lower bound on $\tan\beta$. This
bound is 
sensitive to $A_0$ since $A_0$ enters in the L-R mixing in the stop 
(mass)$^2$ matrix and affects the values of the stop masses. We find for $m_h > 111$ GeV 
(i.e. the 114 GeV experimental bound), that $\tan\beta  > 7$ for $A_0 = 0$, and
$\tan\beta > 5$ for $A_0 = -4m_{1/2}$. At 
higher $m_h$ the bound on $\tan \beta$ is more restrictive. Thus for $m_h > 117$ GeV
(corresponding to an experimental 120 GeV bound), 
one has $\tan\beta > 15$ for $A_0= 0$, and $\tan\beta > 10$ for $A_0 = - 4
m_{1/2}$. As the 
Higgs mass increases, the bound on $\tan\beta$ increases. As discussed in
\cite{b19a,b12,b19,b18}, 
for large $\tan\beta$, the relic density constraints leave only 
co-annihilation regions possible, and these are very 
sensitive to the value of $A_0$. Fig. 1 exhibits the allowed regions  in the 
$m_0 - m_{1/2}$ plane for $\tan\beta =40$,  $m_h  > 111$ GeV for $A_0 =
0,\,-2m_{1/2}$,  
and $4m_{1/2}$ (from bottom to top). The corridors terminate at low $m_{1/2}$
due to 
the $b \rightarrow s\gamma$  and $m_h$ constraints. Without the $a_{\mu}$
constraint, the 
corridors would extend up to the end of the parameter space ($m_{1/2} = 1$ TeV).
We see also that the relic density constraint effectively determines $m_0$ in 
terms of $m_{1/2}$ in this region. The lower bound of Eq. (1), however, cuts
off 
these curves (at the verticle lines) preventing $m_0$ and $m_{1/2}$ from
getting 
too large. Thus for large $\tan\beta$,  the $g_{\mu} -2$ experiment puts a
strong constraint on the SUSY parameter space.

The restriction of the SUSY parameter space by the $a_{\mu}$ constraint affects 
the predicted dark matter detection rates. Thus the exclusion of the large 
$m_0$ and large $m_{1/2}$ domain of Fig. 1 generally raises the lower bounds on
the neutralino-proton cross section. In Fig. 2 we have plotted
$\sigma_{\tilde{\chi}_1^0-p}$  as a 
function of $m_{\tilde{\chi}_1^0}$ for $\tan\beta = 40$ for the allowed
corridors for $A_0  = -2 m_{1/2}, 4 m_{1/2}$ and  0 (bottom to top). The curves
terminate at high $m_{\tilde{\chi}_1^0}$  due 
to the lower bound  on $a_{\mu}$ of Eq. (1). (Note that $m_{\tilde{\chi}_1^0}
\cong 0.4 m_{1/2}$.) Again 
one sees the sensitivity of results to the value of $A_0$, both  for the high 
$m_ {\tilde{\chi}_1^0}$ termination point and for the magnitude of the cross
section. Over the full range one has that $\sigma_{\tilde{\chi}_1^0-p}  > 6
\times 10^{-10}$ pb, and hence should 
generally be accessible to future planned detectors.
\begin{figure}[htb]
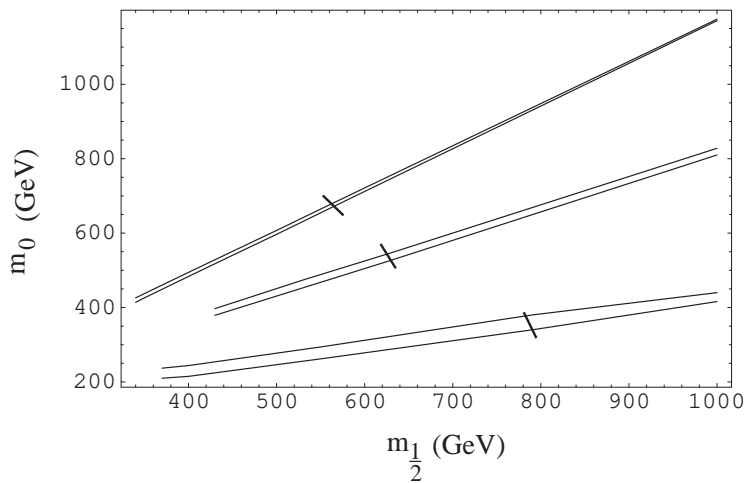

\centerline{ \DESepsf(adhs3.epsf  width 10 cm) }
\caption {\label{fig1} Corridors in the $m_0 - m_{1/2}$ plane allowed by the
relic density 
constraint for $\tan\beta = 40$, $m_h  > 111$ GeV, $\mu > 0$ for $A_0 = 0, 
-2m_{1/2}, 4m_{1/2}$ from bottom to top. The curves terminate at low $m_{1/2}$
due to 
the $b \rightarrow s\gamma$ constraint except for the $A_0 =4m_{1/2}$ which
terminates due to the $m_h$ constraint. The short  lines through the allowed corridors 
represent the
high $m_{1/2}$ termination due to the lower bound on $a_{\mu}$ of Eq. (1).}
\end{figure}

\begin{figure}[htb]
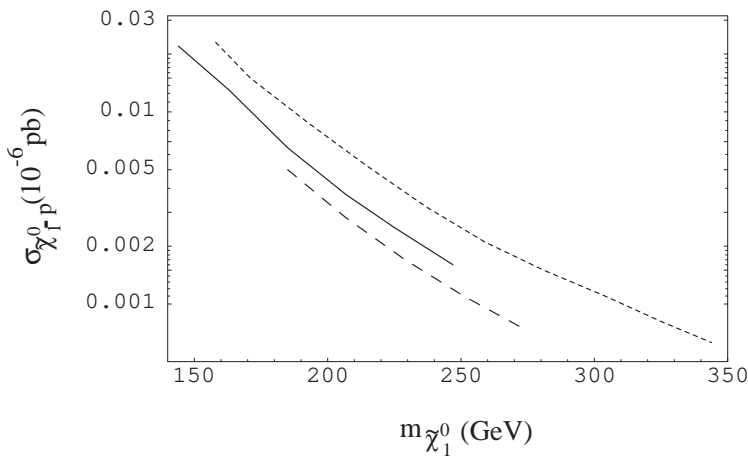

\centerline{ \DESepsf(adhs2.epsf  width 10 cm) }
\caption {\label{fig2} $\sigma_{\tilde{\chi}_1^0-p}$ as a function of the neutralino mass
$m_{\tilde{\chi}_1^0}$ for $\tan\beta = 40$, $\mu > 0$ for $A_0 = -2 m_{1/2}, 4
m_{1/2}, 0$ from bottom to top. The curves terminate 
at small $m_{\tilde{\chi}_1^0}$  due to the $b \rightarrow s\gamma$ constraint
for $A_0 = 0$ and $- 2 m_{1/2}$ and 
due to the Higgs mass bound ($m_h > 111$ GeV) for $A_0 = 4 m_{1/2}$. The curves 
terminate at large $m_{\tilde{\chi}_1^0}$ due to the lower bound on $a_{\mu}$
of Eq. (1).}
\end{figure}

If we reduce $\tan\beta$ , one might expect the minimum value of
$\sigma_{\tilde{\chi}_1^0-p}$ to significantly 
decrease. However, the $a_{\mu}$ bound then becomes more constraining, 
eliminating more and more of the high $m_{1/2}$, high $m_0$ region. This is
shown 
in Fig. 3 where the minimum value of $\sigma_{\tilde{\chi}_1^0-p}$ is plotted
as a function of $m_{\tilde{\chi}_1^0}$, for $\tan\beta = 10$, $\mu > 0$, $m_h >
111$ GeV, for $A_0 = - 4 m_{1/2}$ (lower 
curve), $A_0 = 0$ (upper curve). The $A_0 = 0$ curve terminates at low
$m_{\tilde{\chi}_1^0}$ due to 
the Higgs mass bound, while the $A = -4 m_{1/2}$ terminates due to the $b
\rightarrow s\gamma$ constraint. The termination at high $m_{\tilde{\chi}_1^0}$
is due to the $a_{\mu}$ lower 
bound of Eq. (1). We see that the parameter space is now quite restricted, 
and so even though $\tan\beta$ is quite reduced, we  find
$\sigma_{\tilde{\chi}_1^0-p} > 4 \times 10^{-10}$ pb.  The co-annihilation
region begins at $m_{\tilde{\chi}_1^0} \stackrel{>}{\sim} 140$ GeV, and so 
the earlier part of these curves lie in the non co-annihilation domain.

\begin{figure}[htb]
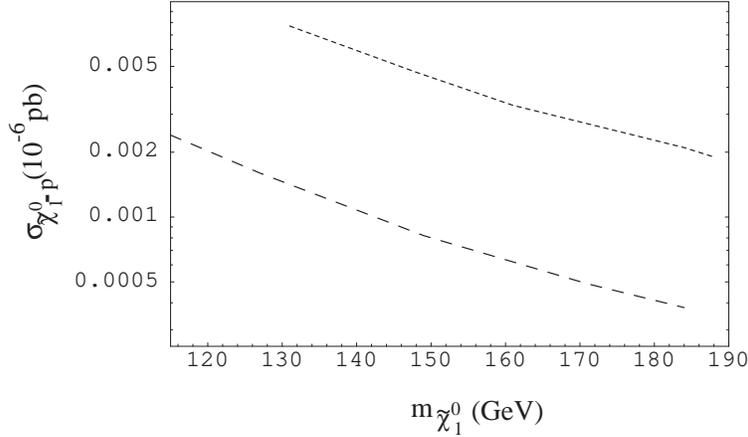

\centerline{ \DESepsf(adhs1.epsf  width 10 cm) }
\caption {\label{fig3} $\sigma_{\tilde{\chi}_1^0-p}$ as a function of $m_{\tilde{\chi}_1^0}$
for $\tan\beta = 10$, $\mu > 0$, $m_h > 111$ 
GeV for $A_0 = 0$ (upper curve), $A_0 = -4 m_{1/2}$ (lower curve). The
termination 
at low $m_{\tilde{\chi}_1^0}$ is due to the $m_h$ bound for $A_0 = 0$, and the
$b \rightarrow  s\gamma$ bound for 
$A_0 = -4 m_{1/2}$. The termination at high $m_{\tilde{\chi}_1^0}$ is due to
the lower bound on $a_{\mu}$ 
of Eq. (1).}
\end{figure}

If we raise $m_h$ and require $m_h > 117$ GeV ( corresponding to an 
experimental bound of 120 GeV), then $m_h$ controls the termination of the 
curves at low $m_{\tilde{\chi}_1^0}$. Thus for  $\tan\beta  = 40$, the curves
of Fig. 2 start at 
$m_{\tilde{\chi}_1^0} = 200$ GeV for $A_0 = -2 m_{1/2}$, at
$m_{\tilde{\chi}_1^0} = 215$ GeV for $A_0 = 0$, and at $m_{\tilde{\chi}_1^0} =
246$ 
GeV for $A_0 =4 m_{1/2}$ (i. e. the $A_0 = 4 m_{1/2}$ curve is almost completely 
eliminated by the $m_h$ constraint). One has thus only a narrow range of 
allowed $m_{\tilde{\chi}_1^0}$ . The allowed range becomes even narrower with
decreasing 
$\tan\beta$, and the entire parameter space is eliminated when $\tan\beta= 10$.

We turn next to consider  non-universal soft breaking models with 
non-universal masses at $M_G$ in the third generation squarks and sleptons 
and in the Higgs masses:
\begin{eqnarray} 
m_{H_{1}}^{\ 2}&=&m_{0}^{2}(1+\delta_{1}); 
\quad m_{H_{2}}^{\ 2}=m_{0}^{2}(1+ \delta_{2});\nonumber \\ m_{q_{L}}^{\
2}&=&m_{0}^{2}(1+\delta_{3}); \quad m_{t_{R}}^{\ 2}=m_{0}^{2}(1+\delta_{4});
\quad m_{\tau_{R}}^{\ 2}=m_{0}^{2}(1+\delta_{5});  \nonumber \\ m_{b_{R}}^{\
2}&=&m_{0}^{2}(1+\delta_{6}); \quad m_{l_{L}}^{\ 2}=m_{0}^{2}(1+\delta_{7}).
\end{eqnarray}
Here $\tilde q_L = (\tilde{t}_L,\tilde{b}_L)$ squarks, $\tilde l_L = (\tilde{\nu}_{\tau},
\tilde{\tau}_L)$ sleptons, etc. 
and  we assume $-1 < \delta_i < +1$. As discussed in \cite{b12}, the value of 
$\mu$  significantly controls both the relic density and
$\sigma_{\tilde{\chi}_1^0-p}$, and one 
may understand qualitatively how $\mu$ varies from its analytic expression which is 
valid for low 
and intermediate $\tan\beta$:
\begin{eqnarray}
\mu^2&=&{t^2\over{t^2-1}}\left[({{1-3 D_0}\over 2}+{1\over
t^2})+{{1-D_0}\over2}(\delta_3+\delta_4)\right. \nonumber \\ 
&-&\left.{{1+D_0}\over2}\delta_2+{\delta_1\over
t^2}\right]m_0^2+{\rm {universal\,parts\,+\,loop \, corrections}}. 
\end{eqnarray}
Here $t = \tan\beta$, and $D_0 \cong 1 - (m_t/200\, {\rm GeV}\sin\beta)^2 \cong
0.25$. One sees 
that the universal $m_0^2$ term is quite small, and one can easily choose the 
$\delta_i$ to make the coefficient of $m_0^2$ negative. A reduction of $\mu^2$ 
increases the higgsino content of the neutralino, and thus increases the 
$\tilde{\chi}_1^0-\tilde{\chi}_1^0- Z$ coupling. In \cite{b12}, it was shown that
this allowed the opening of a 
new region of allowed relic density at high $m_{1/2}$ and high $\tan\beta$. We 
consider first the simple case where only $\delta_2$ is non zero and choose 
$\delta_2 = 1$. Fig. 4 shows $\sigma_{\tilde{\chi}_1^0-p}$ as a function of
$m_{1/2}$ for this case when 
$\tan\beta = 40$, $\mu > 0$, $m_h > 111$ GeV and $A_0 = m_{1/2}$. The lower
line 
corresponds to the usual stau-neutralino co-annihilation corridor. The 
upper dashed curves show the new allowed band arising from increased early 
universe  annihilation through the $Z$ s-channel pole. It is quite broad and 
has a large scattering cross section.  The curves terminate at low $m_{1/2}$
due 
to the $b \rightarrow s\gamma$ constraint, and we have terminated the curves at
the high 
end when $m_0$ or $m_{1/2}$ exceed 1 TeV. The vertical lines are the high
$m_{1/2}$ 
endpoints due to the lower bound on $a_{\mu}$ of Eq. (1). One sees that the 
parameter space is significantly reduced, though there is still a large 
$Z$-channel band remaining. Increasing $m_h$ increases the lower bound of
$m_{1/2}$. 
For $m_h > 117$ GeV we find the co-annihilation (solid line) now begins at 
$m_{1/2} = 510$ GeV, and the $Z$ channel band begins at 500 GeV due to the $m_h$ 
constraint, leaving a sharply reduced region of parameter space.

\begin{figure}[htb]
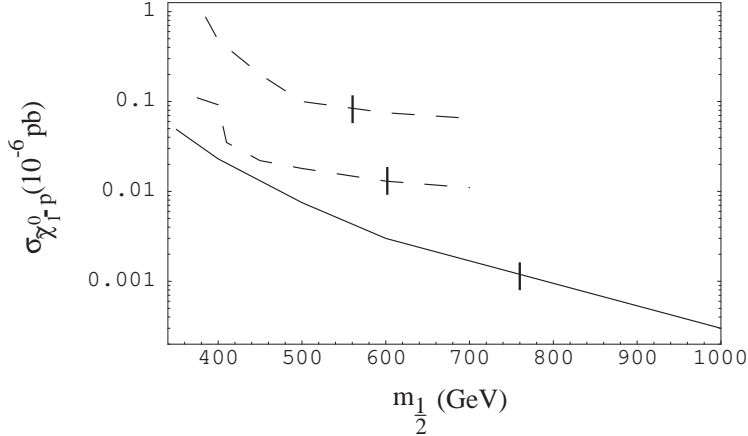

\centerline{ \DESepsf(adhs4.epsf  width 10 cm) }
\caption {\label{fig4} $\sigma_{\tilde{\chi}_1^0-p}$ as a function of $m_{1/2}$
($m_{\tilde{\chi}_1^0} \stackrel{\sim}{=} 0.4 m_{1/2}$) for $\tan\beta = 40$, 
$\mu >0$, $m_h > 111$ GeV, $A_0 = m_{1/2}$ for $\delta_2 = 1$. The lower curve
is for the $\tilde{\tau}_1-\tilde{\chi}_1^0$ co-annihilation channel, and the
dashed band is for the $Z$ s-channel 
annihilation allowed by non-universal soft breaking. The curves terminate 
at low $m_{1/2}$ due to the $b \rightarrow s\gamma$ constraint. The vertical
lines show the 
termination at high $m_{1/2}$ due to the lower bound on $a_{\mu}$ of Eq. (1).}
\end{figure}

A second example of new non-universal effects is furnished by choosing 
$\delta_{10}$ ( $= \delta_3 = \delta_4 = \delta_5$) to be non-zero (as might be
the 
case for an $SU(5)$ or $SO(10)$ model). We consider here $\delta_{10} = - 0.7$. In 
this case~\cite{b12} the $\tilde{\tau}_1-\tilde{\chi}_1^0$ co-annihilation corridor occurs
at a much higher 
value of $m_0$ than in the universal case (i. e. for $m_0 = 600-800$ GeV), and
is 
somewhat broadened.   The $Z$ channel band lies above it and is considerably 
broader. In Fig. 5 we have plotted $\sigma_{\tilde{\chi}_1^0-p}$ as a function
of $m_{1/2}$ for the 
lower side of the co-annihilation corridor (lower curve) and for the upper 
side of the $Z$ channel band (upper curve)  for $\tan\beta = 40$, $\mu >0$, $A_0
= m_{1/2}$ 
and $m_h > 111$ GeV. (Note that while the $Z$ channel lies at a higher $m_0$ in
the 
$m_0 - m_{1/2}$ plane than the co-annihilation corridor, the cross section is 
still larger since $\mu^2$ is reduced.) The curves terminate at the left due 
to the $b \rightarrow s\gamma$ constraint. The vertical lines show the
termination at 
high $m_{1/2}$ due to the lower bound on $a_{\mu}$, significantly shrinking the 
allowed parameter space. For $m_h > 117$ GeV, the Higgs mass governs the 
termination at low $m_{1/2}$, and the co-annihilation (lower curve) now begins 
at $m_{1/2} = 515$ GeV, and the $Z$ channel (upper curve) begins at $m_{1/2} =
520$ GeV.
\begin{figure}[htb]
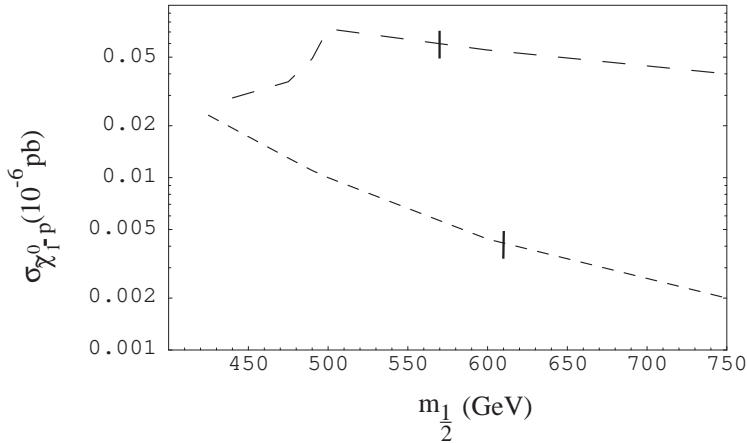

\centerline{ \DESepsf(adhs5.epsf  width 10 cm) }
\caption {\label{fig5} $\sigma_{\tilde{\chi}_1^0-p}$ as a function of $m_{1/2}$ for $\tan\beta =
40$, $\mu >0$, $A_0 = m_{1/2}$ 
and $m_h  > 111$ GeV. The lower curve is for the bottom of the
$\tilde{\tau}_1-\tilde{\chi}_1^0$ 
co-annihilation corridor, and the upper curve is for the top of the $Z$ 
channel band. The termination at low $m_{1/2}$ is due to the $b \rightarrow
s\gamma$ 
constraint, and the vertical lines are the upper bound on  $m_{1/2}$ due to the 
lower bound of $a_{\mu}$ of Eq. (1).}
\end{figure}

The above discussion shows that for SUGRA models, and particularly for 
mSUGRA,  the $a_{\mu}$ data, when combined with the $m_h$, 
$b \rightarrow s\gamma$ and relic 
density constraints have begun to greatly limit the SUSY parameter space. 
Thus the $m_h$ and $b \rightarrow s\gamma$ constraints determine a lower bound 
on $m_{1/2}$ and hence an upper bound on $a_{\mu}^{\rm SUGRA}$, 
while the experimental lower bound on $a_{\mu}$ determines an upper bound on
$m_{1/2}$. The  
combined $a_{\mu}$ and $m_h$ bound puts lower bound on $\tan\beta$ for a given value of 
$A_0$. This can be seen most clearly in Fig. 6, where the mSUGRA contribution 
to $a_{\mu}$ is plotted as a function of $m_{1/2}$
 for $A_0 = 0$, $\tan\beta  = 10$ (lower 
curve), $\tan\beta  = 30$ (middle 
curve)and $\tan\beta = 40$ (upper curve). Further, most of the allowed $m_{1/2}$ 
region lies in the the $\tilde\tau_1-\tilde\chi^0_1$ co-annihilation domain 
( $m_{1/2}\stackrel{>}{\sim} 350$ 
GeV), and so from Fig. 1 one can see that $m_0$ is approximately determined in terms of 
$m_{1/2}$.  In Fig.6, the $m_h$ bound determines the lower limit on $m_{1/2}$ for
$\tan\beta$=10, while $b\rightarrow s\gamma$ determines it for $\tan\beta=40$. Both are
equally constraining for $\tan\beta=$30. If we consider the 90\% C. L. bound 
($a_{\mu} > 21\times10^{-10}$)\cite{b20}), one finds 
for $A_0 = 0$ that $\tan\beta \geq 10$, and for $\tan\beta \leq 40$ that $m_{1/2} = (290 - 550)$ 
GeV, and $m_0 = (70 - 300)$ GeV. This greatly constrains SUSY particle 
spectrum expected at accelerators, as can be seen in Table 1. Thus at the $90\%$
C.L. bound on $a_{\mu}$ the 
tri-lepton signal will be unobservable at the Tevatron Run II since $\tan\beta$ 
and $m_{1/2}$ are relatively large \cite{b21}, and the other SUSY particles are also 
beyond its reach, except for the light Higgs, provided $m_h\stackrel{<}{\sim}130$ GeV
\cite{b22}. (One
would need to triple the Tevatron's energy to see a significant part of the 
SUSY mass spectrum.)   Only the $\tilde\tau_1$ and $\tilde e_1$ would possibly be within the 
reach of a 500 GeV NLC (and very marginally the $\tilde\chi^{\pm}_1$), 
while all the SUSY particles would be accessible to the LHC. The Brookhaven E821 experiment 
has a great deal more data that can reduce the error by a factor of about 
2. When analysed, this would greatly narrow the predictions  made here.

One of the interesting features of Fig. 6 is that mSUGRA can no longer 
accommodate large values of $a_{\mu}^{\rm SUGRA}$. If the full E821 data should 
require a value significantly larger than $40\times10^{-10}$, this would be a 
signal for the existance of non-universal soft breaking. From Eq.(3) one 
sees that one can increase $a_{\mu}$ by reducing $\mu$, and from Eq. (5) this might 
be accomplished by non-universal soft breaking of the scalar masses (and 
also from non-universal gaugino masses at $M_G$.) Thus the $g_{\mu} - 2$ 
experiment may give us significant insight into the nature of physics 
beyond the GUT scale.

\begin{figure}[htb]
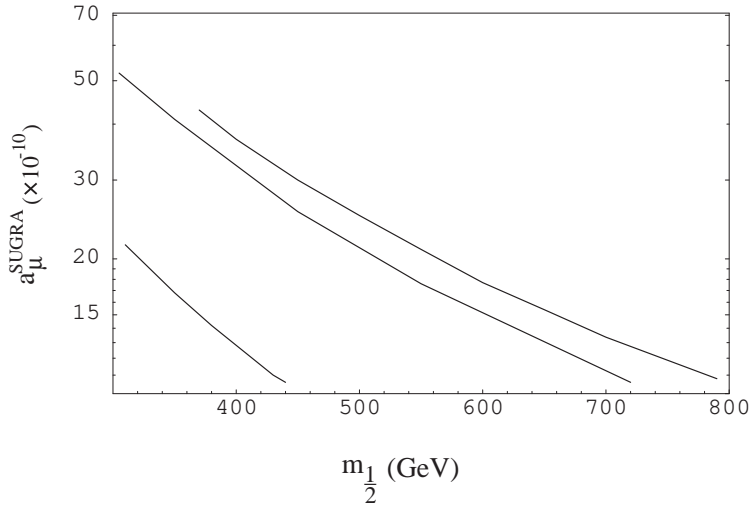

\centerline{ \DESepsf(adhs6.epsf  width 10 cm) }
\caption {\label{fig6}mSUGRA contribution to $a_{\mu}$ as a function of $m_{1/2}$ for $A_0$ = 0,
$\mu > 
0$, for  $\tan\beta = 10$, 30 and 40 (bottom to top). }
\end{figure}

This work was supported in part by National Science Foundation grant No.
PHY-0070964. 
\bigskip

Table 1. Allowed ranges for SUSY masses in GeV for mSUGRA assuming 90\% C. L. 
for  $a_{\mu}$ for $A_0=0$. The lower value of $m_{\tilde t_1}$can be reduced to 240 GeV by
changing $A_0$ to -$4m_{1/2}$. The other masses are not sensitive to $A_0$.
\begin{center}
 \begin{tabular}{|c|c|c|c|c|c|c|}  \hline
$ \tilde\chi^0_1$ &        $\tilde\chi_1^{\pm}$&$\tilde g$ &        $\tilde\tau_1$&
       $\tilde  e_1$&         $\tilde  u_1$&         $\tilde  t_1$\\
  \hline
(123-237)&(230-451)&(740-1350)&(134-264)&(145-366)&(660-1220)&(500-940)\\\hline
\end{tabular}
\end{center}


\begin{thebibliography}{99}
\bibitem{b1} H.N. Brown et.al., Muon (g-2) Collaboration, hep-ex/0102017.
\bibitem{b2} P. Fayet, in \emph{Unification of the Fundamental Particle
Interactions}, edited by S. Ferrara, J. Ellis, and P. van Nieuwenhuizen (Plenum,
New York, 1980); J. A. Grifols and A. Mendez, \Journal{\PRD}{26}{1809}{1982}; J.
Ellis, J. Hagelin and D.V. Nanopoulos, \Journal{\PLB}{116}{283}{1982}; R. Barbieri
and L. Maiani, \Journal{\PLB}{117}{203}{1982}.
\bibitem{b3} S. Ferrara and E. Remiddi, \Journal{\PLB}{53}{347}{1974}.
\bibitem{b4} A.H. Chamseddine, R. Arnowitt and P. Nath,
\Journal{\PRL}{49}{970}{1982}; R. Barbieri, S. Ferrara and C.A. Savoy,
\Journal{\PLB}{119}{343}{1982}; L. Hall, J. Lykken and S. Weinberg,
\Journal{\PRD}{27}{2359}{1983}; P. Nath, R. Arnowitt and A.H. Chamseddine,
\Journal{\NPB}{227}{121}{1983}. 
\bibitem{b5} D.A. Kosower, L.M. Krauss and N. Sakai,
\Journal{\PLB}{133}{305}{1983}.
\bibitem{b6} T.C. Yuan, R. Arnowitt, A.H. Chamseddine and P. Nath,
\Journal{\ZPC}{26}{407}{1984}.
\bibitem{b7} U. Chattopadhyay and P. Nath,
\Journal{\PRD}{53}{1648}{1996}; T. Moroi, \Journal{\PRD}{53}{6565}{1996};
Erratum, ibid, {\bf 56}, 4424 (1997).
M. Carena, G.F. Giudice and C.E.M. Wagner,
\Journal{\PLB}{390}{234}{1997}; T. Goto, Y. Okada and Y. Shimizu, hep-ph/9908499;
T. Blazek, hep-ph/9912460; G.C. Cho, K. Hagiwara and
M. Hayakawa, \Journal{\PLB}{478}{231}{2000}; T. Ibrahim and P. Nath,
\Journal{\PRD}{62}{015004}{2000}.
\bibitem{b8} A. Czarnecki and W.J. Marciano, hep-ph/0010194.
\bibitem{b9} J. Ellis, S. Kelley and D.V. Nanopoulos, \Journal{\PLB}{249}{441}{1990};
 U. Amaldi, W de Boer and H. Furstenau, \Journal{\PLB}{260}{447}{1991}.
\bibitem{b10} J.L. Lopez, D.V. Nanopoulos and X. Wang,
\Journal{\PRD}{49}{366}{1994}.
\bibitem{b11}G. Degrassi, P. Gambino and G. Giudice,
\Journal{JHEP}{0012}{009}{2000}; M. Carena, D. Garcia, U. Nierste and C. Wagner,
\Journal{\PLB}{499}{141}{2001}.
\bibitem{b12} R. Arnowitt, B. Dutta and Y. Santoso, hep-ph/0102181.
\bibitem{a}M. Drees, Y. G. Kim, T. Kobayashi
 and M. Nojiri, hep-ph/0011359. 
\bibitem{b}J. Feng and K. Matchev, hep-ph/0102146. 
\bibitem{b16}U. Chattopadhyay and P. Nath, hep-ph/0102157. 
\bibitem{d}S. Komine, T. Moroi and M. Yamaguchi, hep-ph/0102204. 
\bibitem{b13}J. Ellis, G. Ganis, D.V. Nanopoulos and K.A. Olive, hep-ph/0009355. 
\bibitem{b14} L. Ibanez and C. Lopez, \Journal{\NPB}{233}{511}{1984};
V. Barger, M.S. Berger and P. Ohmann, \Journal{\PRD}{49}{4908}{1994}.
\bibitem{b15} U. Chattopadhyay and P. Nath, \Journal{\PRD}{53}{1648}{1996}.
\bibitem{b17} J. Ellis, A. Ferstl and K.A. Olive,
\Journal{\PLB}{481}{304}{2000}; hep-ph/0007113.
\bibitem{b18} R. Arnowitt, B. Dutta and Y. Santoso, hep-ph/0010244; hep-ph/0101020.
\bibitem{b19a} M. Gomez and J. Vergados, hep-ph/0012020,
 M. Gomez, G. Lazarides and C. Pallis, 
\Journal{\PRD}{61}{123512}{2000}. 
\bibitem{b19} J. Ellis, T. Falk, G. Ganis, K.A.
Olive and M. Srednicki, hep-ph/0102098.
\bibitem{b20} A. Czarnecki and W.J. Marciano, hep-ph/0102122.
\bibitem{b21} V. Barger and C. Kao,  \Journal{\PRD}{60}{115015}{1999}; 
E. Accomando, R. Arnowitt and B. Dutta, \Journal{\PLB}{475}{176}{2000}.
\bibitem{b22}M. Carena et.al., ``{\it{Report of the Tevatron Higgs Working Group}}", 
hep-ph/0010338.
\end{thebibliography}
\end{document}